\documentclass{article}

\usepackage{arxiv}
\usepackage{textcomp}

\usepackage{graphicx}
\usepackage{amsmath}
\usepackage[version=4]{mhchem}
\usepackage{siunitx}
\usepackage{longtable,tabularx}
\usepackage{url}
\usepackage{amsmath}
\usepackage{bm}
\usepackage{enumitem}
\usepackage{float}
\usepackage{natbib}
\usepackage{caption}
\usepackage{subcaption}
\usepackage{array}
\usepackage{amsfonts}
\usepackage[dvipsnames]{xcolor}
\usepackage{booktabs}
\usepackage{nomencl}
\makenomenclature
\usepackage{soul}
\usepackage{arydshln}
\usepackage{placeins}

\usepackage{algorithm}
\usepackage[noend]{algpseudocode}

\usepackage{geometry}
\setcitestyle{numbers} 
\begin{document}
\title{Learning Generative Models for Climbing Aircraft from Radar Data}


\author{
    Nick Pepper \\
	The Alan Turing Institute\\
	The British Library\\
	London, UK \\
	\texttt{npepper@turing.ac.uk} \\
    \And
    Marc Thomas  \\
	NATS\\
	Whitely, Fareham, UK \\
 	\texttt{marc.thomas@nats.co.uk} \\
}

\maketitle

\begin{abstract}
Accurate trajectory prediction (TP) for climbing aircraft is hampered by the presence of epistemic uncertainties concerning aircraft operation, which can lead to significant misspecification between predicted and observed trajectories. This paper proposes a generative model for climbing aircraft in which the standard Base of Aircraft Data (BADA) model is enriched by a functional correction to the thrust that is learned from data. The method offers three features: predictions of the arrival time with 26.7\% less error when compared to BADA; generated trajectories that are realistic when compared to test data; and a means of computing confidence bounds for minimal computational cost.



\end{abstract}

%

\section{Introduction}

Modern Air Traffic Management (ATM) faces the challenge of increasing airspace capacity to meet the growing demand for aviation, while ensuring accurate and safe aircraft operations that meet climate targets \cite{NATS2022, UK_gov}. Within ATM, TP plays an important role in the safe management of airspace, informing decision-making by predicting aircraft trajectories and detecting potential conflicts \cite{bada1, bada2}. However, the task of planning to ensure safe separation is hampered by discrepancies between observed and predicted aircraft trajectories. These discrepancies arise due to the presence of significant epistemic uncertainties, in addition to mis-specification in the TP model.

Sources of this epistemic uncertainty include: uncertainty in forecast wind speed and direction; unknown aircraft speed profile; thrust setting of the aircraft engine; and aircraft mass \cite{Lygeros}. {Aircraft mass and the speed profile are key parameters for predicting climbs}, however, these quantities {can be commercially sensitive and are not routinely shared} with ATM. Current TP methods such as the widely used base of aircraft data (BADA) model \cite{nuic2010bada, nuic2010user} are deterministic. The BADA model is a total energy model conditioned by parameters such as the aircraft mass; speed profile; wind and atmospheric conditions; and mode of flight. The BADA model parameters are calibrated against reference aircraft performance parameters \cite{nuic2010bada}. On the one hand, this allows the model to be generalised to airspace across the world. On the other hand, aircraft on specific routes may follow local procedures or airline operating procedures that result in behaviour very different to what the globally calibrated model expects \cite{Ben2010TP}. 

\begin{figure}[htb!]
\centering
\includegraphics[scale=0.7]{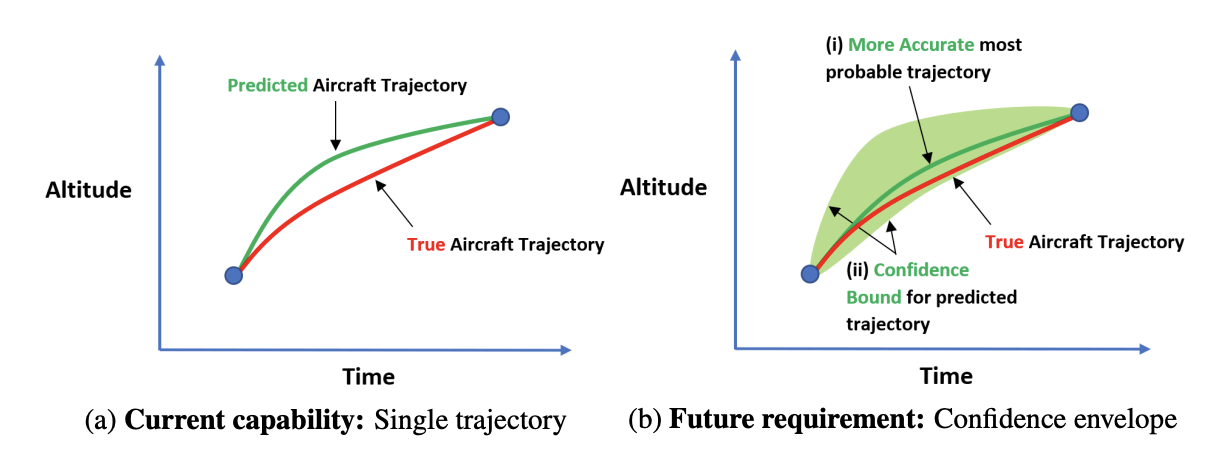}
\caption{(a) schematic illustrating the prediction of the vertical profile of a trajectory using the current paradigm of TP methods. (b) schematic illustrating the desired features of next-generation TP models.}
\label{fig:traj_pred}
\end{figure}

{The general requirements of next-generation TP models are illustrated in Figure \ref{fig:traj_pred}: the model returns a most probable trajectory, learned from situationally relevant data, that is more accurate than existing deterministic models. In addition, the desired model returns a well calibrated confidence bound, that facilitates more accurate prediction tools and more efficient use of available airspace.} 

Two main strategies have been explored within the TP literature for augmenting TP models with data: one strategy uses trajectory data to learn an entirely new TP model altogether using machine learning techniques. A prominent example of this strategy are the methods developed as part of the SESAR DART project, for instance Hernández et al compared Random Forest and Gradient Boosting algorithms \cite{hernandez2018data}, while Fernández et al presented a Hidden Markov Model (HMM) based method for TP \cite{fernandez2017dart}. In the wider literature, Machine Learning methods such as Convolutional and Recurrent Neural Networks \cite{SHAFIENYA2022103878,tran2022aircraft, Bayes+weather}, LSTMS \cite{N-Incept, LSTM, xu2021multi,  shi2018lstm, LSTM2, LSTM3}, Generalized Linear Models \cite{de2013machine}, HMMs \cite{pred_analytics} and clustering approaches \cite{murcca2020data} have all been used for data-driven TP. An advantage of this approach is that there is flexibility in the model to fit the data, however, a drawback is that it is more challenging to enforce physics-based constraints on the modelled trajectories, and to ensure that all predicted trajectories are plausible in the context of ATM. For instance, plausible trajectories for climbing or descending aircraft would be expected to change altitude monotonically \cite{pepper_funcGP} and to follow standard flight procedures, which may not always be the case for purely data-driven methods \cite{Jung}. 

An alternative method which can overcome this drawback is a hybrid strategy, where the physics-based models are retained, but corrections to the default models are inferred from data using nearest neighbours \cite{History, solina_data}, neural networks \cite{alligier2015machine, hybrid_fuel}, gradient boosting machines \cite{alligier2018learning}, Markov Chain Monte Carlo \cite{SUN2020391} or linear and polynomial regression \cite{ALLIGIER201345, YANTO2018574}. This provides a more flexible approach to informing and improving TP models with data that retains the ATM driven constraints and modelling.


This paper proposes a probabilistic, hybrid model for climbing aircraft. An effective thrust function is learned from data to fit the BADA model to specific scenarios. In this paper we focus on aircraft climbing within an interval of altitudes in upper airspace. The model has three features, which will be demonstrated: 
\begin{itemize}
    \item Improved mean prediction of aircraft climbs, compared to BADA using nominal parameters
    \item The model is generative and can be used to generate realistic synthetic trajectories
    \item Upper and lower confidence bounds on the performance of a climbing aircraft that can be computed for minimal additional computational cost 
\end{itemize}

The remainder of this paper is structured as follows: in Section \ref{sec:method} the probabilistic method is outlined in detail. Section \ref{sec:data} describes the real-world dataset, harvested from aircraft surveillance data, that was used to train and then validate the model. Finally, Section \ref{sec:res} contrasts the performance of the model against BADA using nominal parameters for a test dataset.

\section{Methodology}\label{sec:method}
\subsection{Non-identifiability of BADA parameters for climbing aircraft}
BADA is a physics-based model describing the flight mechanics of an aircraft. The total energy equation is defined in BADA as: 
\begin{align}
    (T_{HR}-D)V_{TAS}=m\bigg(g_0\frac{dh}{dt}+V_{TAS}\frac{dV_{TAS}}{dt}\bigg), 
    \label{eq:tot_en}
\end{align}
where $T_{HR}$ represents the aircraft thrust parallel to the velocity vector; $D$ the drag; $V_{TAS}$ the true airspeed of the aircraft; $m$ the aircraft mass; $h$ the geodetic altitude of the aircraft; and $g_0$ the gravitational constant. The LHS of the equation represents the excess power available to an aircraft to either climb or increase its speed. The two terms on the RHS represent the rate of change of gravitational potential energy and kinetic energy respectively. 
The BADA manual describes how \eqref{eq:tot_en} may be rearranged to give an equation for the rate of climb/descent, $ROCD$ \cite{nuic2010user}:
\begin{align}
    ROCD=\frac{T-\Delta T}{T}\bigg[\frac{(T_{HR}-D)V_{TAS}}{mg_0}\bigg]f(M),
    \label{eq:rocd}
\end{align}
where $T$ represents the atmospheric temperature of the International Standard Atmosphere (ISA) model; $\Delta T$ is a correction factor that allows the ISA model to be adjusted if meteorological data is available; and $f(M)$ is the energy share factor (ESF) which reflects how much power is allocated to climbing as opposed to acceleration, expressed as a function of the Mach number, $M$. The functional form of $f(M)$ is dependent on the flight condition. 
Using equations 3.6-1, 3.6-2, and 3.6-5 in the BADA handbook \cite{nuic2010user}, \eqref{eq:rocd} can be rearranged, yielding the expression:
\begin{align}
    \frac{2g_0^2c_{D2}}{\text{cos}^2(\phi)\rho V_{TAS}S}m^2+
    \bigg[\frac{ROCD\times T}{f(M)(T-\Delta T)}\bigg]mg_0+
    \frac{1}{2}c_{D0}\rho V_{TAS}^3S-T_{HR}V_{TAS}=0,
    \label{eq:no_id}
\end{align}
where $c_{D0}$ and $c_{D2}$ are drag coefficients, $\rho$ the air density (from the ISA model), $S$ the reference wing surface area, and $\phi$ is the banking angle. In \eqref{eq:no_id} four parameters are unknown: $ROCD$, $m$, $V_{TAS}$, and $T_{HR}$. To be able to learn corrections from data, $ROCD$ is observed from radar data ($ROCD^*$ is used to denote an observation of the $ROCD$) and nominal values are used for $V_{TAS}$ and $m$, allowing an effective thrust term, $\hat{T}_{HR}$, to be fitted. This term is a function of geodetic altitude, $h$, and acts as a functional correction to the BADA equations, representing a combination of the thrust of the aircraft and corrections to misspecification of other BADA parameters. \eqref{eq:no_id} can be rearranged to express the fitted effective thrust as a function of $ROCD^*$ and terms in the BADA model with nominal values. The following subsection describes the generative model used to model this functional correction and how it is fitted to data. 

\subsection{Generative model for corrections to the thrust profile}
This section describes a generative model for the functional form of the fitted thrust. $\hat{T}_{HR}(h)$ is modelled as a stochastic process, discretized over a grid of $n_g$ geodetic altitudes. {Fitting $\hat{T}_{HR}(h)$ }to a dataset of $n_f$ flights and interpolating to this grid provides $n_f$ realisations of this stochastic process. Functional Principal Component Analysis (fPCA) is used to represent the variation in $\hat{T}_{HR}(h)$ as a weighted sum of $n$ orthonormal basis functions \cite{ramsay2013functional}:
\begin{align}
    \hat{T}_{HR}(h_j)=\mu(h_j)+\sum_{i=1}^{n}w_{i}\phi_i(h_j),\;j=1:n_g,
    \label{eq:fpca}
\end{align}
where $\mu(h_j)=\mathbb{E}(\hat{T}_{HR}(h_j))$ represents the mean of the stochastic process and $\phi$ the discretized orthonormal basis functions, satisfying:
\begin{align}
    \int \phi_i(h)\phi_j(h) dh=\delta_{ij}.
\end{align}
Functional PCA follows similar principals to Principal Component Analysis (PCA), however, while PCA determines the eigenvalues and eigenvectors of a covariance matrix, estimated from samples, in fPCA a set of eigenvalues and eigenfunctions is sought for a covariance function. 
Having determined the mean function and orthonormal basis functions, a least squares fitting is performed to fit each of the $n_f$ flights in the dataset. The weights associated with one flight are denoted as the point $\boldsymbol{w}=[w_1,\dots, w_n]^\top$ in a space of fPCA weights, $\mathbb{W}$, with $\boldsymbol{w}\in\mathbb{W}\subseteq\Re^{n}$. Repeating the process of fitting the trajectory data for each of the $n_f$ trajectories yields the set $W=[\boldsymbol{w}^{(1)},\dots, \boldsymbol{w}^{(n_f)}]$. 

The joint density of the fPCA coefficients, $f_w(\boldsymbol{w})$, is approximated using a multi-variate normal distribution $f_w(\boldsymbol{w})\approx N(\boldsymbol{\mu}_w, \Sigma_w)$, where $\boldsymbol{\mu}_w\in \mathbb{W}$ and $\Sigma_w\in\Re^{n\times n}$ represent the distribution mean and covariance matrix respectively. The subscript `$w$' is used to distinguish this mean from that in \eqref{eq:fpca}. 

Three specific advantages of this model are: firstly, fitting BADA to data allows us to improve its mean accuracy and tune the model to specific locations and scenarios for which we have data. Secondly, the representation can be used as a generative model for aircraft climbs, $\hat{T}_{HR}$ can be generated by sampling weights from this distribution and using \eqref{eq:fpca} within BADA to compute a synthetic climb profile. Lastly, it provides a convenient way of defining a data-driven upper and lower bound on the performance of climbing aircraft, which is discussed in the next section.


\subsection{Optimised uncertainty bounds for climbing aircraft}\label{sec:method_bnds}
{TP models are used in ATM to place conservative bounds around predicted trajectories, in order to assist controllers with planning safe routes for aircraft. Ideally, these bounds should be accurate, informed by real-world data, and inexpensive to compute i.e. analytic bounds that can be computed without sampling}. Inspired by recent developments in optimised model uncertainty in Neural Networks \cite{heiss2021nomu}, and by geometrical arguments for computing worse-case failure probabilities in reliability analysis \cite{PEPPER2022108635}, this section introduces an analytical method for computing upper and lower confidence bounds on the hybrid model described above. 

Level sets of the likelihood can be described as $n$-dimensional ellipsoids in $\mathbb{W}$, the fPCA weight space. Given that the covariance matrix, $\Sigma_w$, is diagonal due to the orthonormal basis of the fPCA modes, the axis lengths of these ellipsoids are determined by the square root of the diagonal elements of $\Sigma_w$, with the ellipsoids being axis aligned. Choosing the 95\% confidence level results in a confidence ellipsoid enclosing points that satisfy: 
\begin{align}
    (\boldsymbol{w}-\boldsymbol{\mu}_w)^\top\Sigma_w^{-1}(\boldsymbol{w}-\boldsymbol{\mu}_w)\leq\chi^2_{0.95}, 
\end{align}
where $\chi^2_{0.95}$ represents an evaluation of the $\chi^2$ distribution with $n$-degrees of freedom at the $0.95$ confidence level. A search is performed to identify the points within the confidence ellipsoid that maximises and minimises the thrust at each point in the grid of geodetic altitudes. These points are denoted $\hat{\boldsymbol{w}}_u^{(k)}$ and $\hat{\boldsymbol{w}}^{(k)}_l$ with $k=[1:n_g]$. 

By making the substitution $a_i=\phi_i(h_k)$, \eqref{eq:fpca}, evaluated at the $k$\textsuperscript{th} geodetic altitude, can be written as:
\begin{align}
    \boldsymbol{a}^\top \boldsymbol{w}=\text{const.}, 
    \label{eq:hyp}
\end{align}
where $\boldsymbol{a}=[a_1,\dots,a_n]^\top$ collects the evaluations of the fPCA basis functions. \eqref{eq:hyp} effectively defines a family of hyperplanes in $\mathbb{W}$, all with unit normal $\hat{\boldsymbol{a}}=\boldsymbol{a}/\|\boldsymbol{a}\|^2$. $\hat{\boldsymbol{w}}_u^{(k)}$ and $\hat{\boldsymbol{w}}_l^{(k)}$ correspond to locations on the surface of the confidence ellipsoid that are tangential to this family of hyperplanes. It can be shown that these locations correspond to: 
\begin{align}
    \hat{\boldsymbol{w}}_l^{(k)}=\boldsymbol{\mu}_w-\sqrt{\chi^2_{0.95}}\Big(\boldsymbol{c}\circ\hat{\boldsymbol{n}}\Big) \\
    \hat{\boldsymbol{w}}_u^{(k)}=\boldsymbol{\mu}_w+\sqrt{\chi^2_{0.95}}\Big(\boldsymbol{c}\circ\hat{\boldsymbol{n}}\Big), \nonumber
\end{align}
where $\boldsymbol{c}=\text{diag}(\Sigma_w)^{\frac{1}{2}}$ and $\boldsymbol{n}=\boldsymbol{a}\circ\boldsymbol{c}$, with unit normal $\hat{\boldsymbol{n}}=\boldsymbol{n}/\|\boldsymbol{n}\|^2$.

Having determined the sets of locations within the confidence ellipsoid $\hat{\boldsymbol{w}}_u^{(k)}$ and $\hat{\boldsymbol{w}}_l^{(k)}$ with $k=[1:n_g]$, an upper and lower bound for the climbing performance can be obtained by interpolating to obtain a best-case and worst-case thrust profile and by evaluating BADA once for each profile. Therefore, this enables a computationally efficient analytic calculation of these bounds.



\section{Data Preparation}\label{sec:data}

\begin{figure}
\begin{center}
    	\includegraphics[width=0.49\columnwidth]{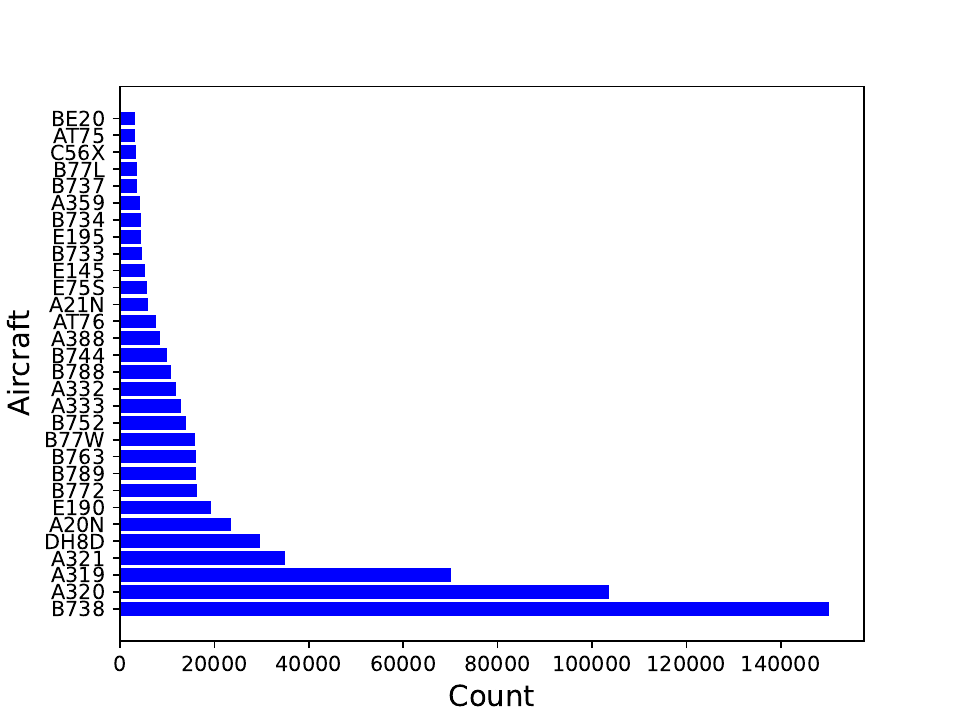}
     	\includegraphics[width=0.49\columnwidth]{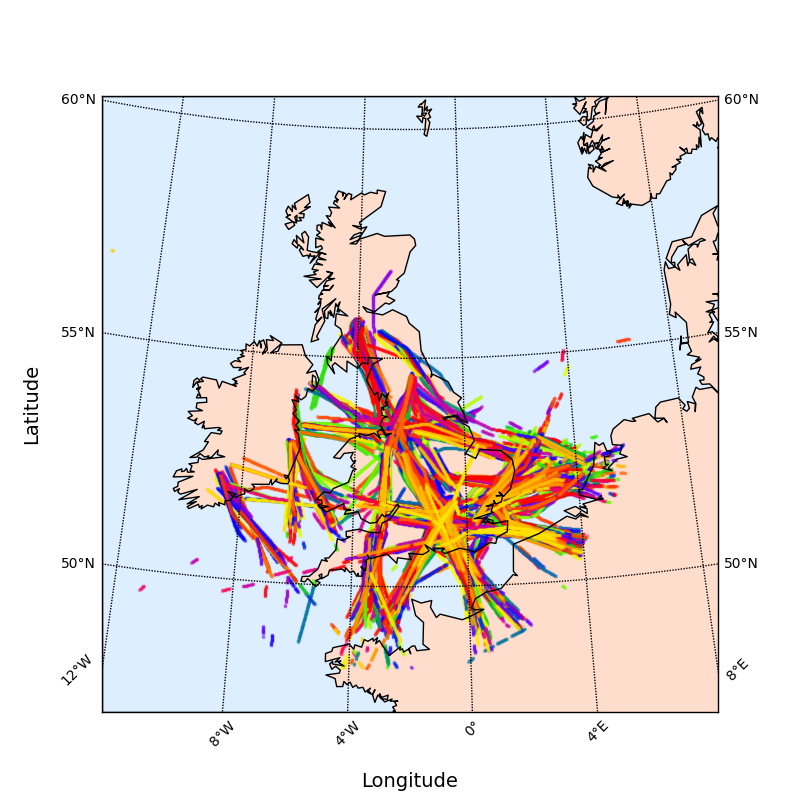}
    \caption{Frequency at which each aircraft type occurs in the dataset. For ease of display, displayed aircraft types were filtered to those with more than 3000 appearances in the dataset (left). Lateral positions of the B738 radar blips in the dataset with $ROCD\geq500\text{ft/min}$ (right).}
    \label{fig:ac_count}
    \end{center}
\end{figure}

The method described in Section \ref{sec:method} for learning a functional correction to the thrust term of the BADA equations from data was tested using a dataset of real-world trajectory data. This dataset comprised Mode S radar surveillance data from 707,236 flights, collected between July and September 2019 across the southern portion of UK airspace (London FIR \cite{nats_fir}). The left panel of Figure \ref{fig:ac_count} displays the frequency at which aircraft types occurs within the dataset. For clarity, only those aircraft types that occur more than 3000 times in the dataset are displayed. The dataset is dominated by B738 and A320 aircraft. 

A generative model is developed for frequently occurring aircraft types, flying between flight levels 150 and 325. These flight levels were selected as they are in the upper airspace, where trajectories are less likely to be affected by local operational procedures. As a pre-processing step, the dataset was filtered to radar blips with $ROCD\geq500\text{ft/min}$ that belonged to trajectories that climbed through this interval. A 66:33 train:test split was performed on this filtered dataset. The right panel of Figure \ref{fig:ac_count} illustrates the lateral position of the B738 radar blips, colored by the flight they are associated with.

A point-wise approach was then followed to fit $\hat{T}_{HR}$ to the training data. The $\hat{T}_{HR}$ samples associated with individual trajectories were interpolated to an equally spaced grid of $n_g=100$ points within the flight level interval, producing a dataset of functional data for each aircraft type on which fPCA was performed. An analysis of the explained variance using the kneedle algorithm \cite{satopaa2011finding} was used to determine the number of retained principal components. {Typically 2-4 components were retained for each aircraft type}.

\begin{figure}
\begin{center}
    \includegraphics[width=0.49\columnwidth]{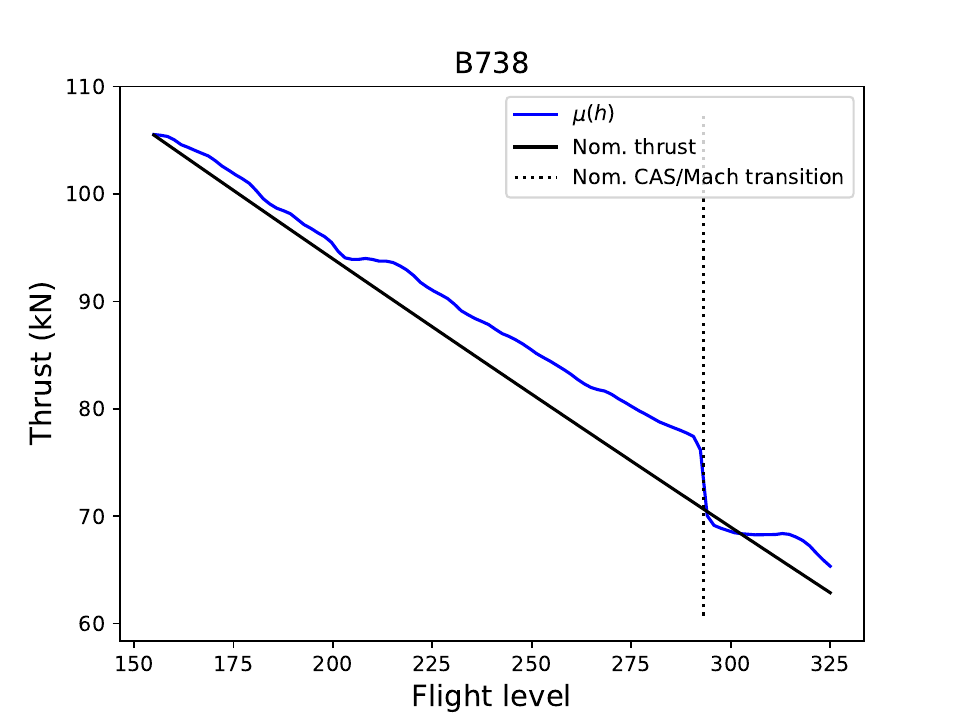}
    \includegraphics[width=0.49\columnwidth]{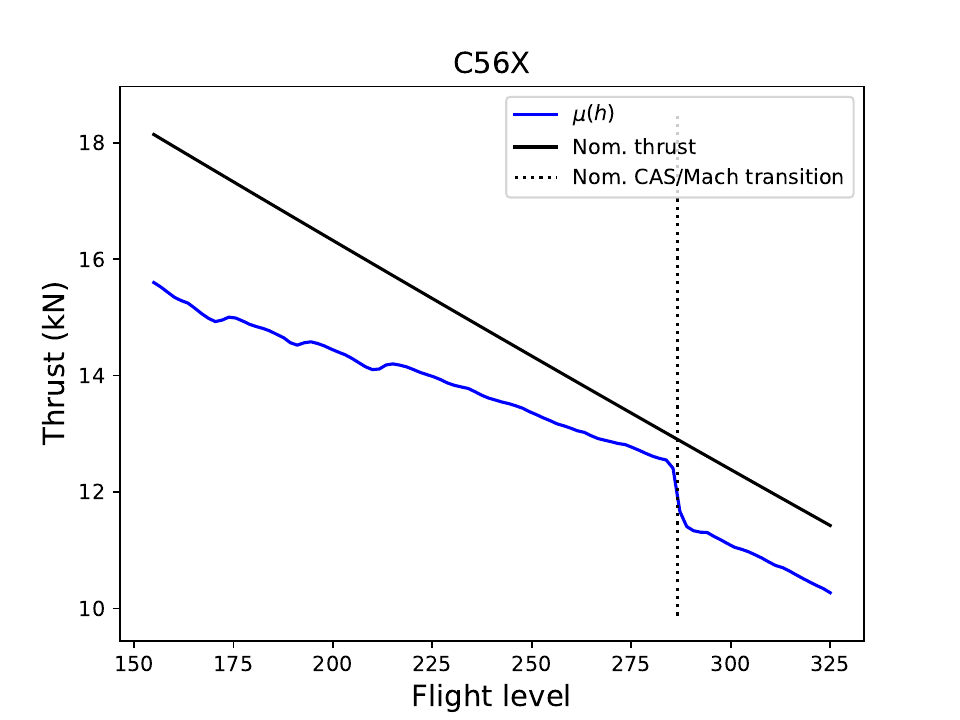}
    \caption{Fitted mean thrust as a function of flight level for three aircraft types. Vertical lines indicate the CAS-Mach transition altitude using the BADA speed schedule.}
    \label{fig:pseudo_th}
    \end{center}
\end{figure}

Figure \ref{fig:pseudo_th} illustrates the fPCA mean, $\mu(h)$, plotted against flight level for the B738 and C56X aircraft types. The B738 was chosen as an example of frequently occurring passenger jets, while the Cessna C56X is a smaller corporate jet chosen as a comparison with markedly different performance characteristics. In both sub-plots the fitted thrust profile is roughly linear. Compared to the nominal thrust profile, the gradient of the fitted B738 is steeper, while in the case of the C56X $\mu(h)$ is shifted to be roughly $2kN$ lower than the nominal thrust profile. $\mu(h)$ for both the B738 and C56X has a discontinuity around the altitude of the CAS-Mach transition, indicated by a vertical dotted line. This altitude is calculated in BADA 3.16 using the nominal speed schedule for each aircraft type, suggesting that the fitted $\hat{T}_{HR}$ is compensating for misspecification in the nominal BADA speed profiles and/or mass parameter, for the specific routes in the dataset. 

Having determined the mean function and fPCA modes, the weights associated with each trajectory in the training dataset were determined through a least squares fitting. These weights were collected as the set $W$, to which a multi-variate normal distribution was fitted. The fitting process was repeated independently for each aircraft type in the training dataset. 

\section{Results}\label{sec:res}
Having fitted the hybrid model for a range of aircraft types, the proposed method was evaluated using the test dataset. Three criteria were chosen for the assessment: how well the model was able to improve on the mean error of the BADA model using nominal parameters; how well the upper and lower confidence bounds, generated using the 95\% confidence interval of the model weights, contained the test data; and how well synthetic trajectories generated by the model reflected the distribution of trajectories in the test dataset. 

The left panels of the subplots in Figure \ref{fig:results1} illustrate sampled thrust profiles generated by the model, their mean, and upper and lower bounds determined using the method in Section \ref{sec:method_bnds} at the 95\% confidence level. For comparison, the nominal BADA thrust profile is plotted. The green line, $T_{th, min}$, indicates the thrust profile corresponding to $ROCD=0$ at each flight level. The centre panel of each subplot compares the trajectories in the test dataset to the mean and bounds of the model, when the thrust profiles are used in BADA. The green line in these plots corresponds to BADA run with nominal parameters. This real world test data can be compared to the trajectories corresponding to the generated thrust profiles and the model bounds, displayed in the right hand panel. For each aircraft type, a set of samples was generated equal to the number of trajectories in the test dataset, generating synthetic climb profiles that appear very similar to the test data. In this section plots were generated for the B738 and A320, as these two aircraft types dominated the dataset, and C56X as an example of a smaller aircraft with markedly different performance characteristics.

\subsection{Mean predictions}
\begin{figure}
\begin{center}
    \includegraphics[width=1.0\columnwidth]{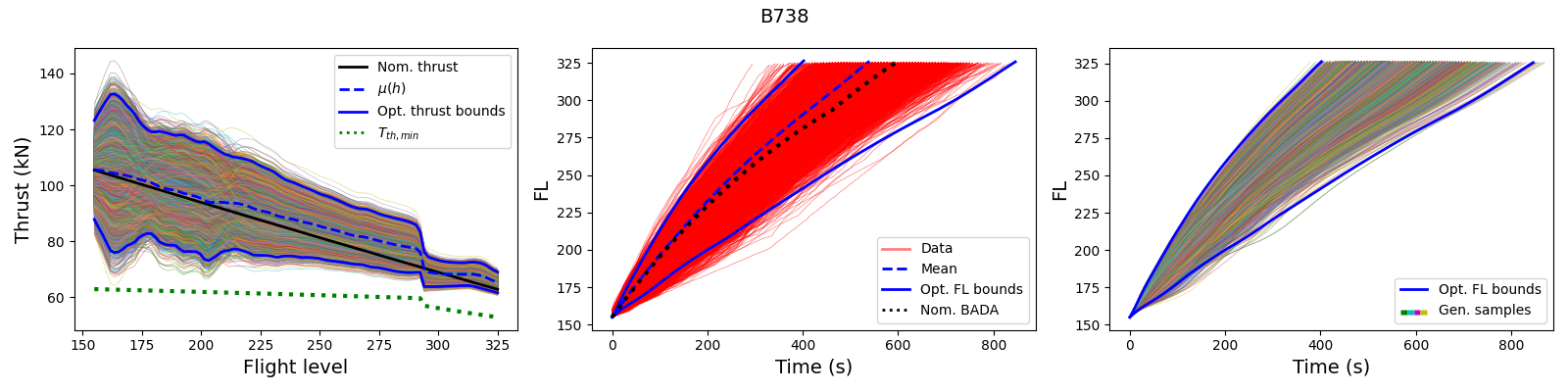}
    \includegraphics[width=1.0\columnwidth]{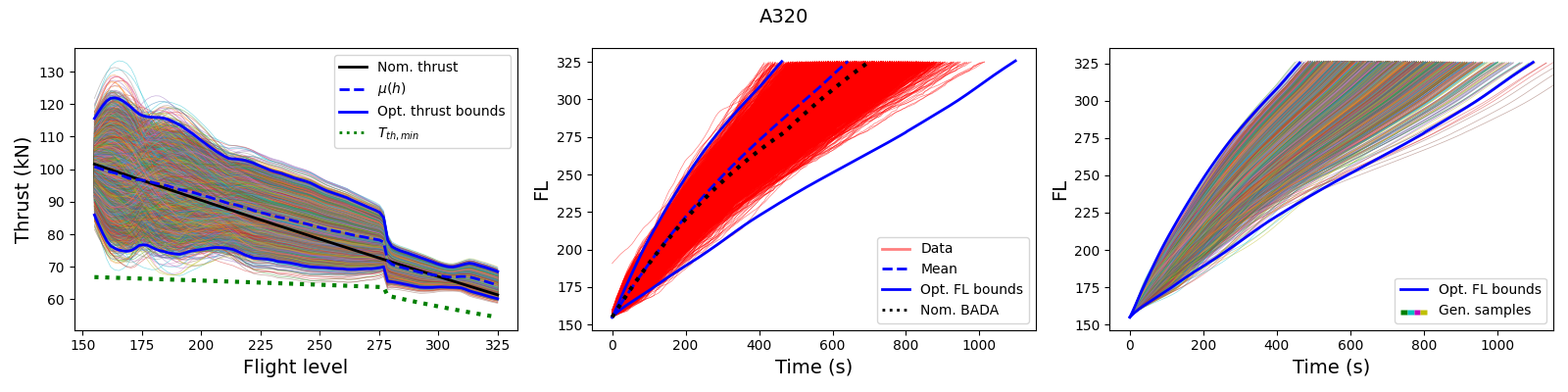}
    \includegraphics[width=1.0\columnwidth]
    {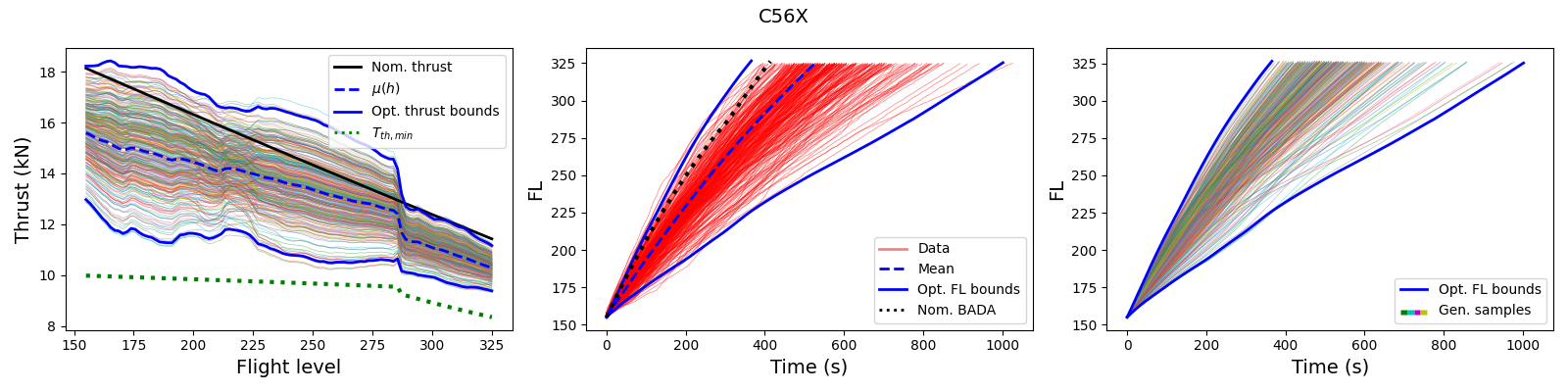}
    \caption{Samples of the thrust function with optimised bounds, compared to the nominal BADA thrust profile (left). Test data compared to the generative model mean and optimised uncertainty bounds and nominal BADA (middle). Sampled trajectories and uncertainty bounds when thrust samples and bounds run through BADA (right).} 
    \label{fig:results1}
    \end{center}
\end{figure}

The top left panel of Figure \ref{fig:results2} is a swarm plot that displays the mean error at two flight levels: 250 (roughly halfway in the interval) and 325 for every aircraft type. The data-driven model in blue can be seen to have a significantly lower mean absolute error (MAE) than the BADA model, in red. The data-driven model significantly improves the MAE for all aircraft types, particularly towards the top of the climbs, this can be seen in the middle panels of Figure \ref{fig:results1}. 

Table 1 displays results for all aircraft types in the dataset. On average, the MAE, across flight levels 250 and 325, for the data-driven model is {26.7\%} lower than the predicted trajectory of BADA using nominal parameters.

\subsection{Accuracy of generative model}
\begin{figure}
\begin{center}
    \includegraphics[width=0.49\columnwidth]{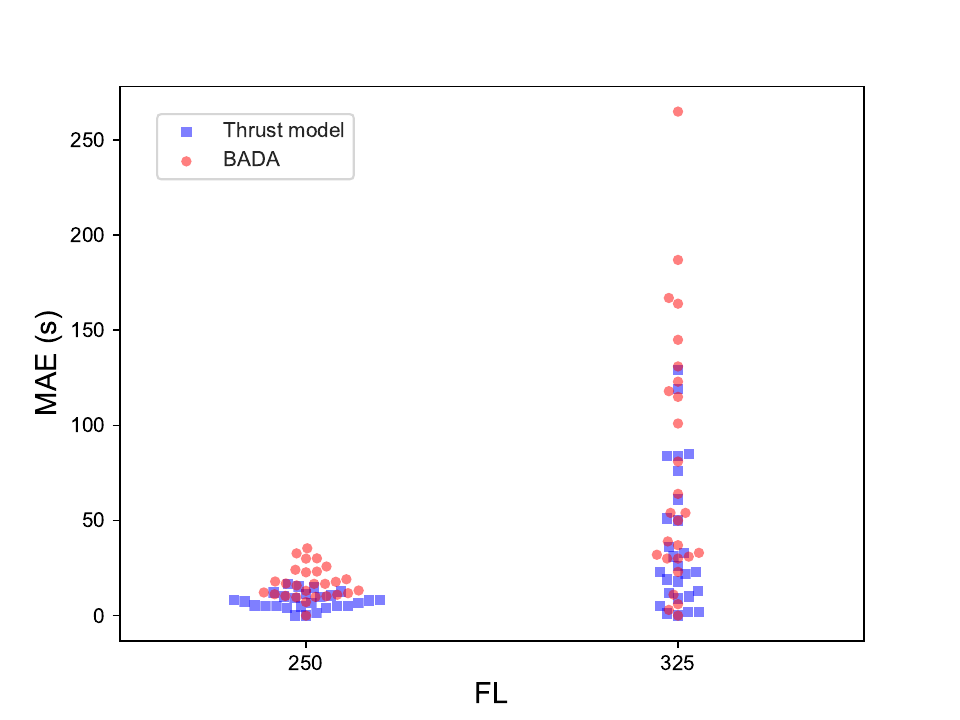}
    \includegraphics[width=0.49\columnwidth]{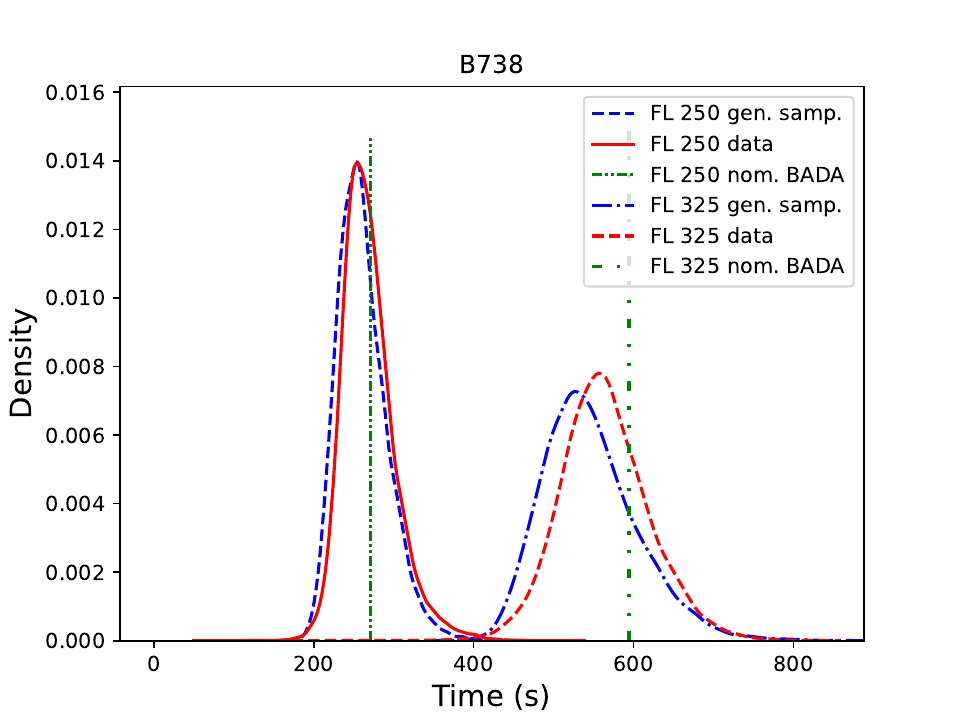}
    \includegraphics[width=0.49\columnwidth]{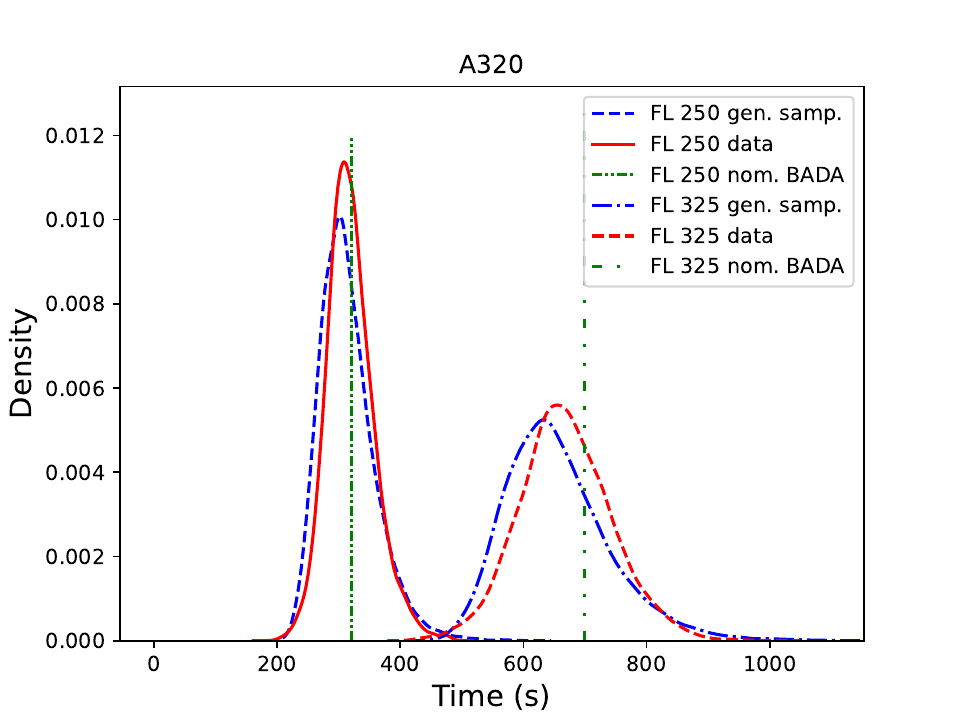}
    \includegraphics[width=0.49\columnwidth]{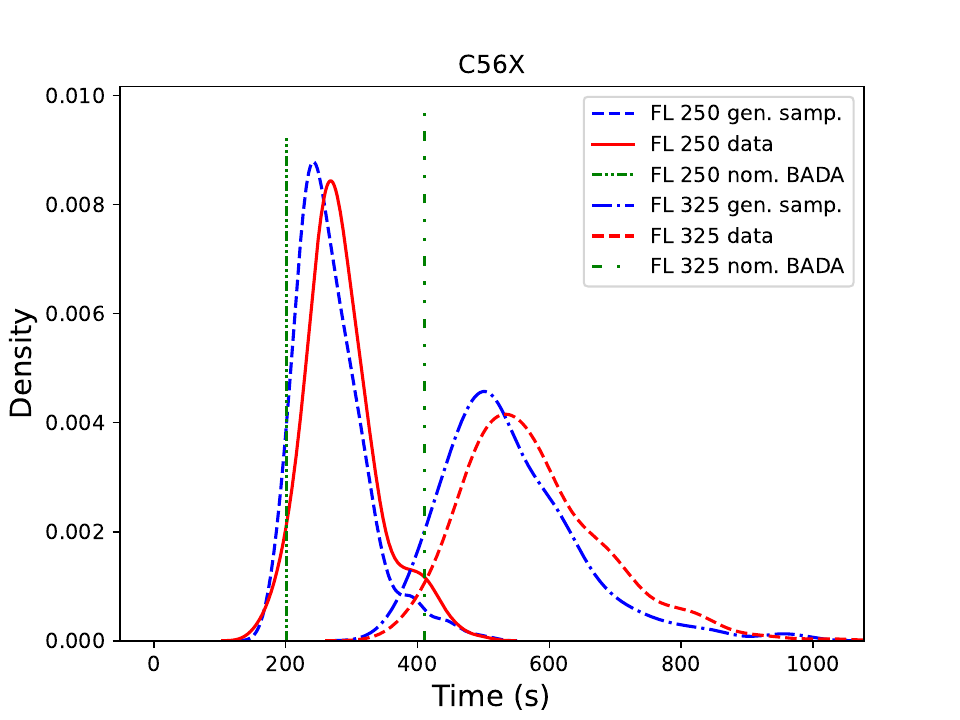}
    \caption{ Swarm plot indicating the MAE at Fl 250 and 325 for the proposed model and BADA (top left). Kernel density estimates of the distribution of arrival times at two flight levels, with the test data (red) compared to trajectories sampled from the generative model (blue).}
    \label{fig:results2}
    \end{center}
\end{figure}

The ability of the model to generate realistic trajectories was tested using the statistical distance between probability distributions for the arrival times at flight levels 250 and 325. Figure \ref{fig:results2} displays kernel density plots of these distributions. For comparison, vertical lines are plotted that represent the arrival time predicted using BADA with nominal parameters (which is deterministic). With the exception of the arrival of the B738 at FL250 in Figure \ref{fig:results2}, the BADA prediction appears to underestimate the arrival time.

The statistical distance between the distributions generated by the model and the test dataset was computed using the KL-divergence. The majority of values were in the range 0.3-0.8, indicating good agreement between the distributions. One of the largest statistical distances was for the B738 at FL250 (3.9), this is plotted in the top left panel of Figure \ref{fig:results2} and nevertheless appears to show reasonable agreement. 

\subsection{Containment of bounds}

Finally, the ability of the proposed method to produce accurate confidence bounds was investigated. Bounds were generated for each aircraft type at the 95\% confidence level, with the proportion of the test dataset lying within the bounds recorded in Table 1. The bounds were found to be highly accurate for most aircraft types: the average coverage of the test data was {97.2\%. For all aircraft coverage in the training dataset was greater than 95\%}.


\FloatBarrier
\section{Conclusions}
A generative model has been proposed that uses a functional correction to the thrust profile of the BADA model. Trained on data harvested from 707,236 flights across UK airspace, a probabilistic generative model was developed for 26 common aircraft types. The models offers three main features: an improved climb trajectory, giving a mean estimate of arrival time at FL325 that was on average {26.7\%} more accurate than BADA run with nominal parameters; the ability to generate realistic synthetic trajectories of climbing aircraft; and a highly efficient analytical method for computing upper and lower bounds on climbing aircraft for a specified confidence level generated on our training dataset, {97.2\%} of the test dataset was within these bounds. Given the importance of TP to supporting decision making in ATM, this is a key feature as it allows the method to generate best and worst case climbing profiles within a given confidence level that are informed by data. {Future work will aim to extend this methodology to other scenarios in ATM, such as modelling descents or trajectories following specific ATM procedures.}

\section*{Appendix}

\begin{table}[h]
\centering
\caption{Performance of the proposed generative model for a range of aircraft types, when compared to unseen test data. Arrows indicate whether larger of smaller values are preferable for each metric.}
\begin{tabular}{ccccccccc}
\toprule
Aircraft type & $n_f$& \multicolumn{4}{c}{Mean error (s) $\downarrow$} & \multicolumn{2}{c}{KL div. $\downarrow$}&
Coverage (\%)\\

& & \multicolumn{2}{c}{FL 250} & \multicolumn{2}{c}{FL 325} & \\
 & & Gen. Model & BADA & Gen. Model & BADA  &FL 250 & FL 325 \\
\midrule

B738 & 36955 &   8.0 &   1.48 & 17.77 &  30.38 & 0.11 & 0.19 & 97.02 \\
A320 & 25460 &  4.92 &   1.53 & 12.16 &  30.97 & 0.11 & 0.14 & 97.88 \\
A319 & 16898 &  3.92 &  12.09 & 13.04 &  39.13 &  0.1 & 0.17 & 97.89 \\
A321 &  8296 &  5.21 &  30.71 & 10.97 &  32.33 & 0.12 & 0.14 & 98.05 \\
A20N &  5918 &  4.95 &  21.66 & 17.93 &  50.39 & 0.09 &  0.2 & 97.55 \\
E190 &  4855 & 10.49 &  19.37 & 24.09 &   3.21 &  0.1 & 0.15 & 97.25 \\
B772 &  3037 &  6.55 &  61.27 &  16.7 & 117.69 & 0.12 & 0.14 &  97.3 \\
B789 &  3242 &   8.1 & 118.95 & 16.69 & 164.48 & 0.07 & 0.09 & 97.08 \\
B763 &  1687 &  15.3 &    5.0 & 29.95 &  30.43 & 0.22 & 0.24 & 97.44 \\
B77W &  1865 &   6.7 &   2.07 &  9.62 &  81.03 & 0.14 & 0.15 & 97.23 \\
B752 &  3146 &  9.63 &  50.47 & 16.89 & 101.28 & 0.18 &  0.2 & 96.42 \\
A333 &  2099 &  4.05 & 128.77 & 13.19 & 265.31 &  0.1 & 0.09 &  96.9 \\
A332 &  1950 &  4.86 &  84.02 &  7.14 & 187.21 & 0.11 & 0.13 & 97.82 \\
B788 &  2120 &  5.11 &  36.15 & 11.78 &   5.63 & 0.05 & 0.06 & 97.44 \\
B744 &  1612 &  7.48 &  84.54 &  9.84 & 131.41 & 0.17 & 0.15 & 96.89 \\
A388 &  1423 &  0.09 &  75.77 & 10.34 &  64.25 & 0.15 &  0.1 & 97.21 \\
A21N &  1655 &  8.09 &  18.27 & 19.11 & 115.07 & 0.13 & 0.17 & 97.97 \\
E75S &   755 &  12.2 &  23.23 & 25.81 &  53.78 & 0.19 & 0.21 & 97.39 \\
E145 &   927 &  9.31 &  51.33 & 22.76 & 144.91 & 0.13 & 0.17 & 96.89 \\
B733 &   582 &  2.15 &  10.23 & 13.62 &  23.37 & 0.12 & 0.21 & 96.87 \\
E195 &   743 &	12.75 & 33.45 &	30.1 & 32.85 & 	0.17 & 	0.23 & 97.14 \\
B734 & 669 & 	14.87 & 13.16 & 32.72 & 53.8 &	0.22 & 0.25 &	96.57 \\
A359 &	720 &	10.09 &	22.88 &	10.2 &	37.06 &	0.18 &	0.16 &	97.01 \\
B737 & 	868 &	11.29 &	8.71 &	23.16 &	11.12 &	0.14 &	0.18 &	96.17 \\
B77L &	249 &	5.0 &	26.4 &	15.86 &	123.37 &	0.18 &	0.22 & 97.54\\
C56X	& 697	&16.59&	84.32&	35.39&	166.79&	0.2&	0.19&	96.3 \\
\bottomrule
\end{tabular}
\label{Table:1}
\end{table}

\section*{Funding Sources}
The work described in this article is primarily funded by the grant “EP/V056522/1 : Advancing Probabilistic Machine Learning to Deliver Safer, More Efficient and Predictable Air Traffic Control” (aka Project Bluebird), an EPSRC Prosperity Partnership between NATS, The Alan Turing Institute, Exeter, and Cambridge.

\section*{Acknowledgments}
The authors would like to thank Shubhani Jain, Jan Povala, Richard Everson, and George {De Ath} for their helpful comments. 

\bibliography{biblo} 
\end{document}